\documentclass[sn-mathphys-num]{sn-jnl}% Math and Physical Sciences Numbered Reference Style
\usepackage{graphicx}%
\usepackage{multirow}%
\usepackage{amsmath,amssymb,amsfonts}%
\usepackage{amsthm}%
\usepackage{mathrsfs}%
\usepackage[title]{appendix}%
\usepackage{xcolor}%
\usepackage{textcomp}%
\usepackage{manyfoot}%
\usepackage{booktabs}%
\usepackage{algorithm}%
\usepackage{algorithmicx}%
\usepackage{algpseudocode}%
\usepackage{listings}%
\theoremstyle{thmstyleone}%
\theoremstyle{thmstyletwo}%

\theoremstyle{thmstylethree}%

\begin{document}

\title[Article Title]{Error-correctable efficient quantum homomorphic encryption using Calderbank–Shor–Steane codes}


\author*[1]{\fnm{IlKwon} \sur{Sohn}}\email{d2estiny@kisti.re.kr}

\author[1]{\fnm{Boseon} \sur{Kim}}\email{boseon12@kisti.re.kr}

\author[1]{\fnm{Kwangil} \sur{Bae}}\email{kibae@kisti.re.kr}

\author[1]{\fnm{Wooyeong} \sur{Song}}\email{wysong@kisti.re.kr}

\author[1]{\fnm{Wonhyuk} \sur{Lee}}\email{livezone@kisti.re.kr}

\affil*[1]{\orgdiv{Quantum Network Research center}, \orgname{Korea Institute of Science and Technology Information}, \orgaddress{\city{Daejeon}, \postcode{34141}, \country{Republic of Korea}}}


\abstract{The integration of quantum error correction codes and homomorphic encryption schemes is essential for achieving fault-tolerant secure cloud quantum computing.
However, owing to the significant overheads associated with these schemes, their efficiency is paramount.
In this study, we develop an efficient quantum homomorphic encryption scheme based on quantum error correction codes that uses a single encoding process to simultaneously perform encryption and encoding.
By using a longer quantum error correction code, both the security and error-correction capabilities of the scheme are improved.
Through comprehensive evaluations, we demonstrate that the proposed scheme is more secure than the conventional permutation-key-based QHE scheme when the number of maximally mixed states is not more than twice the length of the quantum error-correction code.
The proposed scheme offers a more secure and efficient approach to quantum cloud computing, potentially paving the way for more practical and scalable quantum cryptographic protocols.}




\maketitle
\section{Introduction}
Cloud computing, which refers to the use of virtual computing resources over the Internet~\cite{CC09}, has gained popularity owing to the increasing availability of internet connectivity across various devices.
However, as all operations are performed on a server, security issues still persist~\cite{sec12}.
Moreover, the service provider can access the data stored in the cloud, thereby posing the threat of potential personal or sensitive data leaks.
Therefore, the data must be encrypted before being sent to the server.
Standard symmetric key encryption does not fully address the issue of information leakage because it requires the data to be decrypted before executing operations.
These security issues can be solved through homomorphic encryption (HE)~\cite{HES}, which allows complex mathematical operations to be performed on encrypted data without decryption~\cite{HE0, HE1, HE2, HE3, HE4, HE5, HE6, HE7, HE8, HE9, HE10}.

Quantum HE (QHE)~\cite{QHE1, QHE2, QHE3, ou18, lai18, QHE4, QHE5, QHE6, QHE7, ou22} has emerged as a viable solution for securely encrypting quantum information, especially considering that cloud-computing systems are widely employed in current commercial quantum-computing platforms~\cite{QCC}, and large-scale quantum networks are becoming more feasible.
In particular, several studies have investigated using the logical gate construction properties of quantum error-correction codes (QECCs) to implement QHE~\cite{ou18, lai18, ou22}.
These QHE based on QECCs offers the advantage of fault-tolerant quantum computing because the encryption is based on random quantum codes, which simplifies the application of additional codes for error correction.
However, these methods handle HE and error correction separately and require individual encoding for both, leading to inefficient resource usage.
This results in the consumption of considerable resources because a concatenated code structure is required for fault-tolerant universal quantum computation~\cite{KN05, ch17, fo18, tr22, no22}.

To address these challenges, this paper introduces an error-correctable, efficient QHE scheme by encoding with a single quantum error correction code. Our scheme leverages the inherent properties of QECCs to perform encryption and error correction simultaneously, eliminating the need for separate encoding steps and reducing resource consumption. Through detailed numerical analysis, we demonstrate that the proposed scheme  offers higher security than other QHE schemes~\cite{ou18, ou22}, particularly effective for 84.10\% of combinations involving maximally mixed states (MMSs) and QECC lengths of up to 50 qubits. Furthermore, our scheme achieves this improved security without a significant increase in resource requirements, making it a practical and scalable solution for fault-tolerant secure quantum cloud computing.

\section{Permutation-key-based quantum homomorphic encryption supports quantum error correction}
\begin{figure}[t!]
\centering
\includegraphics[width=.5\linewidth]{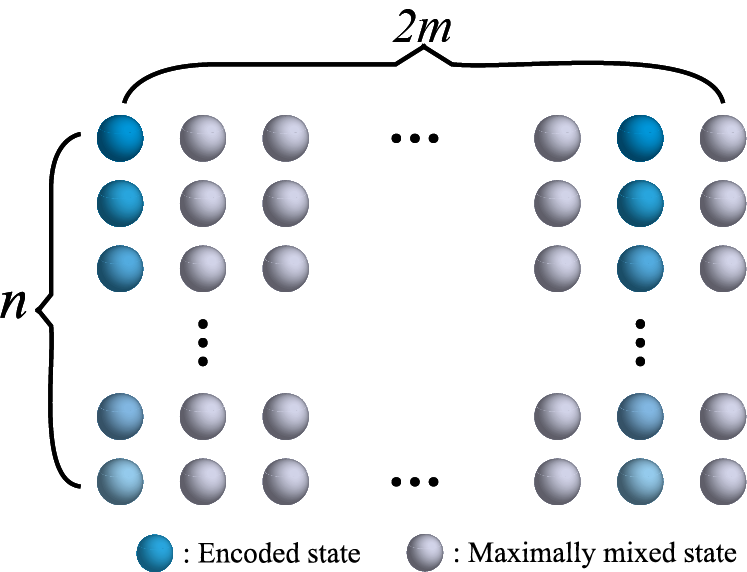}
\caption{\textbf{Schematic of the quantum homomorphic encryption scheme proposed in ~\cite{ou22}}: $2m$ denotes the length of the maximally mixed state columns required for encryption and $n$ denotes the length of the quantum error correction code.
Based on each factor, permutation-key-based quantum homomorphic encryption is performed over two encoding rounds.}
\label{fig:presch}
\end{figure}

This section describes the permutation-key-based QHE, which offers the advantage of simple error correction assisted by the QECC formalism~\cite{ou18, ou22} compared with the random Pauli key-based QHE scheme~\cite{QHE2, ph19}.
Fig.~\ref{fig:presch} shows the final state after a qubit of a message is encrypted using the permutation-key-based QHE scheme proposed in~\cite{ou22}.
The encryption process comprises the following steps:
\begin{enumerate}
    \item[1.] The message is expressed in coordinates starting from the top-left corner as (1,1) and located at position (1,1), whereas the remaining $2mn-1$ qubits are in MMSs, which are represented by gray spheres.
    \item[2.] QECC encoding is applied to the $n$ qubits in the first column.
    \item[3.] The $m$ columns in the front are encoded row-wise into random quantum codes. The blue spheres represent their encoded states.
    \item[4.] The final state is obtained through column-wise permutation of the remaining half of columns and those encoded with random quantum codes.
\end{enumerate}
The QECC used for error correction is concatenated with the one used for encryption and disperses messages into the MMSs.
Thus, $m$ and $n$ should be increased to enhance the error-correction capability and improve security.

Although concatenation is advantageous for suppressing errors in quantum computers~\cite{KN05, CL16} and constructing gate sets for universal quantum computing~\cite{OC14, yd16}, the resource overhead increases exponentially as the concatenation level increases~\cite{KN05}.
This overhead is particularly significant in non-transversal gate operations that require several magic states~\cite{bv05}.
Consequently, various studies have focused on mitigating the concatenation overhead~\cite{no22, We15, so19}.
However, we propose an approach that performs both encoding and encryption in a single encoding step, thereby simultaneously securing error correction and security and lowering the computational overhead by avoiding concatenation.

\section{Error correctable efficient QHE scheme using single encoding}
This section presents an efficient QHE scheme that corrects errors using a single encoding.
This scheme is further elucidated in Section~\ref{DP}, whereas Sections~\ref{tg},~\ref{ntg}, and~\ref{se} elaborate on its support for fault-tolerant universal quantum computing.

\subsection{Permutation-key-based QHE scheme using single encoding}
\label{DP}
This section systematically describes the proposed symmetric QHE scheme that uses the same key for encryption and decryption~\cite{QHE1}.

\subsubsection{Definition} A symmetric quantum homomorphic encryption scheme $\mathcal{H}$:
\begin{enumerate}
    \item[] Key Generating algorithm $KeyGen_{\mathcal{H}}$ is for generating a key $\kappa$;
    \item[] $Encrypt_{\mathcal{H}}$($\mathcal{E}_{\mathcal{H}}$) is the encryption algorithm: $\rho_{\mathcal{H}} = \mathcal{E}_{\mathcal{H}}(\kappa,\rho_m)$;
    \item[] $Decrypt_{\mathcal{H}}$($\mathcal{D}_{\mathcal{H}}$) is the decryption algorithm: $\rho_m = \mathcal{D}_{\mathcal{H}}(\kappa,\rho_{\mathcal{H}})$;
    \item[] $Evaluate_{\mathcal{H}}(\textit{Eval}_{\mathcal{H}})$ is the evaluation algorithm: $\textit{Eval}_{\mathcal{H}}(\kappa, U_{C,\kappa},\rho_{\mathcal{H}})$.
\end{enumerate}
The quantum state used for encryption, i.e., the plaintext, is denoted as $\rho_m$, and the ciphertext is denoted as $\rho_{\mathcal{H}}$.
All evaluation operations are denoted as $U_{C,\kappa}$, representing the general operation $U_{C}$ encrypted using $\kappa$.
$U_{C,\kappa}$ is decomposed and approximated according to the universal quantum-gate set $\{H, T, CNOT\}$.
The result of the evaluation algorithm is equivalent to $\mathcal{E}_{\mathcal{H}}(\kappa, Eval(U_{C},\rho_m))$, where $Eval$ denotes the evaluation process without the QHE.
We demonstrate the proposed scheme based on this definition.

\subsubsection{Proposed QHE scheme}
\begin{figure}[t]
\centering
\includegraphics[width=1\linewidth]{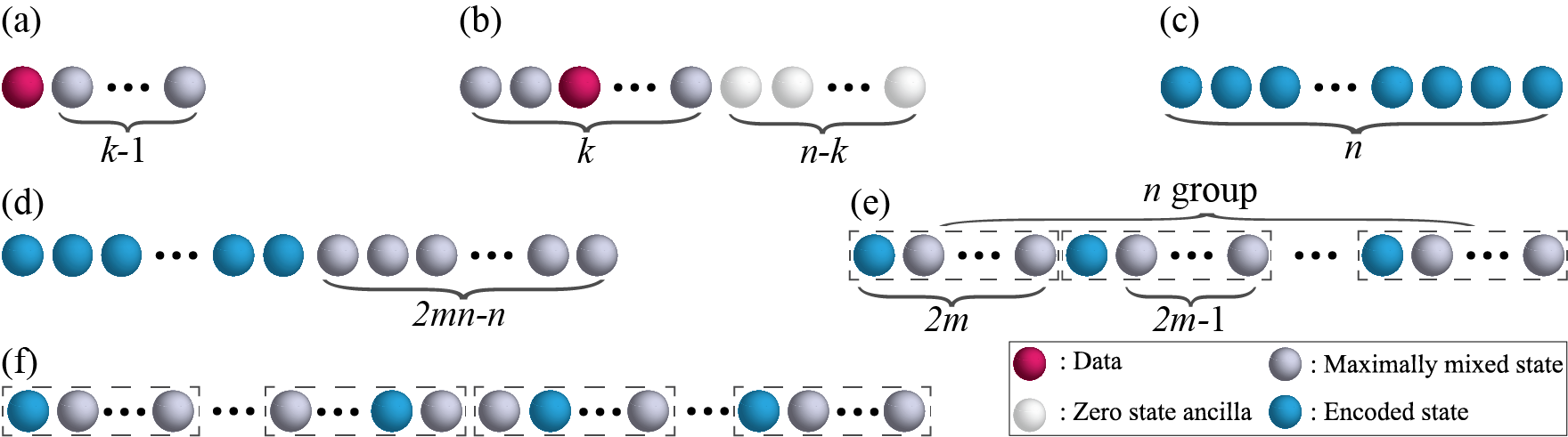}
\caption{\textbf{Schematic picture of the proposed scheme}:
(a) One qubit of a message and $k-1$ maximally mixed states. (b) Addition of $n-k$ zero-state ancilla qubits to implement quantum error-correction codes. (c) $n$ qubit encoded state. (d) Addition of $2mn-n$ maximally mixed states for quantum homomorphic encryption. (e) Grouping one qubit of the encoded state with $2m-1$ maximally mixed states. (f) Random permutation of $2m$ qubits within the group.}
\label{fig:oursch}
\end{figure}

This scheme involves the following processes, as illustrated in Fig.~\ref{fig:oursch}:
\begin{itemize}
  \item[] \textbf{$KeyGen_{\mathcal{H}}$:}
    \begin{itemize}
      \item[] \textbf{Input:} None.
      \item[] \textbf{Processing:} Generate a random bit-string $\kappa=\kappa_1\kappa_2...\kappa_n$, where each $\kappa_i$ is of length $2m$ with Hamming weight 1.
      \item[] \textbf{Output:} The random key $\kappa$.
    \end{itemize}

  \item[] \textbf{$Encrypt_{\mathcal{H}}$:}
    \begin{itemize}
      \item[] \textbf{Input:} The one-qubit message state $\rho_m$, MMSs, zero-state ancilla qubits, an arbitrary permutation operator $P$, a QECC encoding operator $U_E$, and a permutation operator $P_{\kappa}$ according to $\kappa$.
      \item[] \textbf{Processing:} Compute
      \[
      \rho_{\mathcal{H}} = P_{\kappa} [U_E (P \otimes I^{\otimes n-k}) \rho_{m||a} (P^{\dagger} \otimes I^{\otimes n-k}) U^{\dagger}_E \otimes \left( \frac{I}{2} \right)^{\otimes 2mn-n}]P^{\dagger}_{\kappa},
      \]
      where $\rho_{m||a}$ represents $\rho_m$ concatenated with $k-1$ MMSs and $n-k$ zero-state ancilla qubits.

      Through this process, $\rho_m$ is concealed among the $k-1$ MMSs and undergoes QECC encoding alongside $n-k$ zero-state ancilla qubits.
      Then, using $P_{\kappa}$, the $2mn-n$ MMSs are grouped into sets of $2m-1$ qubits, within which the encoded state is interleaved one qubit at a time to complete the encryption.
      \item[] \textbf{Output:} The encrypted state $\rho_{\mathcal{H}}$.
    \end{itemize}

  \item[] \textbf{$Decrypt_{\mathcal{H}}$:}
    \begin{itemize}
      \item[] \textbf{Input:} Encrypted state $\rho_{\mathcal{H}}$ and key $\kappa$.
      \item[] \textbf{Processing:} Compute
      \[
      (P \otimes I^{\otimes n-k}) U_E \mathrm{Tr_{MMS}}[\rho_{\mathcal{H}}] U^{\dagger}_E (P^{\dagger} \otimes I^{\otimes n-k}).
      \]
      In the decryption process, decrypting the parts encrypted via $P_{\kappa}$ can be accomplished without reversing this operation.
      Instead, it is done by partially tracing out the MMSs located in positions where $\kappa$ has zeros, as indicated by $\mathrm{Tr_{MMS}}$.
      \item[] \textbf{Output:} The decrypted message state $\rho_m$.
    \end{itemize}

  \item[] \textbf{$Evaluate_{\mathcal{H}}$:}
    \begin{itemize}
      \item[] \textbf{Input:} Encrypted state $\rho_{\mathcal{H}}$ and a unitary operation $U_{C,\kappa}$.
      \item[] \textbf{Processing:} Compute
      \[
      \rho_O = U_{C,\kappa} \rho_{\mathcal{H}} U^{\dagger}_{C,\kappa},
      \]
      where $U_{C,\kappa}$ is a unitary transformation applied homomorphically and depends on the key.
      \item[] \textbf{Output:} The evaluated state $\rho_O$.
    \end{itemize}
\end{itemize}

\subsubsection{KeyGeneration.} The client generates $\kappa$ used as encryption keys and constructs $P_{\kappa}$.
Even if the specifications of the employed QECC and number of qubits within each group are disclosed, decryption remains theoretically challenging for servers or eavesdroppers.
Subsequently, the client prepares a one-qubit message state $\rho_m$ for computation.
Additionally, the permutation-key $\kappa$ of the proposed scheme with a length of $2mn$ is $n$ times longer than that used in the permutation-key-based QHE scheme proposed in~\cite{ou22} with a length of $2m$.
If the entity delegating the computation to the server differs from that receiving the result, the encryption-key transmission uses $n$-times more data.
However, we primarily assumed scenarios involving a single user.

\subsubsection{Encryption.} First, $\rho_{m||a}$ must be computed for the QECC encoding as follows:
\begin{equation}
\rho_{m||a} = \rho_m \otimes \left(\frac{I}{2}\right)^{\otimes k-1} \otimes |0 \rangle\langle 0|^{\otimes n-k}.
\end{equation}
Subsequently, using the permutation operator $P$, $\rho_m$ is placed at random positions among the MMSs.
Mixing the message qubit at arbitrary positions within $k-1$ MMSs enhances security because it transforms the problem into $\binom{k}{1}$.
However, this is not considered in Section~\ref{SP}.
Furthermore, from the perspective of logical operations, operators that perform the same operation on all the expanded $k$-qubit messages are used for the homomorphic operations.
Therefore, these operations are not affected by the positions of the actual message qubits, as explained in Section~\ref{tg}.

The QECC encoding operator $U_E$ is applied to $\rho_{m||a}$ to generate the encoded state $\rho_E$, thus implementing the $[[n,k,d]]$ error correction code.
However, the proposed scheme does not exploit all QECCs.
Owing to syndrome extraction and non-transversal gate operation, the QECCs are confined to self-dual and doubly even Calderbank–Shor–Steane (CSS) codes~\cite{cal96, st96, st97}, such as the $[[7,1,3]]$ Steane code.
Although applying this code to arbitrary QECCs may be advantageous, transversality is prioritized owing to the quantum-computing overhead, necessitating the use of self-dual and doubly even CSS codes with transverse $H$, $CNOT$, and $S$ gates.

Thus, the encoded state, $\rho_E$, is expressed as
\begin{equation}
\rho_E = U_E (P\otimes I^{\otimes n-k})\left[\rho_{m||a}\right](P^{\dagger}\otimes I^{\otimes n-k}) U^{\dagger}_E,
\end{equation}
where $\rho_{E||a}=\rho_E \otimes \left( \frac{I}{2} \right)^{\otimes 2mn-n}$.
Finally using the previously constructed $P_{\kappa}$ to permute the qubits of $\rho_{E||a}$, the QHE state is obtained as follows:
\begin{equation}
\rho_{\mathcal{H}} = P_{\kappa}\left[\rho_{E||a}\right] P^{\dagger}_{\kappa}.
\end{equation}
The client receives $r$ from the server which is the specific number of required $T$-gates, along with the counts of logical zero and plus states.
Upon receiving this information, the client prepares magic, logical-zero, and logical-plus states using the same method used to generate the message.
These states, along with the encrypted message, $\rho_{\mathcal{H}}$, and gate-operation set, are then transmitted to the server, indicating its readiness to perform cloud-computing tasks.

\subsubsection{Evaluation.}
The server decomposes and approximates the circuit of the desired quantum algorithm into gates using a universal quantum-gate set $\{H, T, CNOT\}$.
Additionally, $r$ which is the total number of $T$-gate operations, required is communicated to the client, which then prepares the appropriate number of magic states.
Further details regarding the universality are provided in Sections~\ref{tg} and~\ref{ntg}.
Upon receiving $\rho_{\mathcal{H}}$, ancilla states, and the gate-operation set from the client, the server performs error correction to rectify any errors that may have occurred during the quantum-state transmission.
Using the logical ancilla states and $\rho_{\mathcal{H}}$ received from the client, the server performs the decomposed and approximated operation $U_C$.
While performing the $T$-gate operations or error correction, which are described in Sections~\ref{ntg} and~\ref{se}, respectively, the server supplements the measurement-output interpretations by communicating classically with the client.
Once all operations are completed, the server obtains the output state $\rho_{O}$, which is defined as
\begin{equation}
\rho_{O}=U_{C,\kappa} \rho_{\mathcal{H}}U_{C,\kappa}^{\dagger}.
\end{equation}
and transmits it to the client.

\subsubsection{Decryption.}
The client does not have to reverse the entire encryption process for decryption because it already knows the identity of each qubit.
Thus, it can obtain the resulting state $\rho_R$ by tracing out all MMSs of $\rho_O$ according to $\kappa$, where
\begin{equation}
\rho_R = \mathrm{Tr_{MMS}}\left[\rho_O\right].
\end{equation}
Finally, the computation result can be obtained through a logical measurement on $\rho_R$~\cite{LM23}.

To analyze the decryption complexity of the proposed scheme, the number of SWAPs required for decryption is $O((r+1) \times nm)$ because the proposed scheme uses only one row of data.
This complexity can then be expressed as $O(r^3 + r^2)$ because $m$ and $n$ are linear in $r$.
Consequently, the proposed scheme is compact for a polynomial number of $T$-gates.
However, as it is based on leveled fully HE (FHE)~\cite{QHE5,lfhe01, lfhe02}, it cannot be bootstrapped or used for computing an arbitrarily large number of $T$-gates.

\subsection{Transversal gate operation}
To evaluate the feasibility of fault-tolerant universal quantum computing, a more detailed discussion beyond the scope of the evaluation is necessary.
From this section onward, an in-depth explanation of performing transversal gate operations, non-transversal $T$-gate operations, and syndrome extraction will be provided for the fault-tolerant universal quantum computing.
\label{tg}
\begin{figure}[t!]
\centering
\includegraphics[width=.85\linewidth]{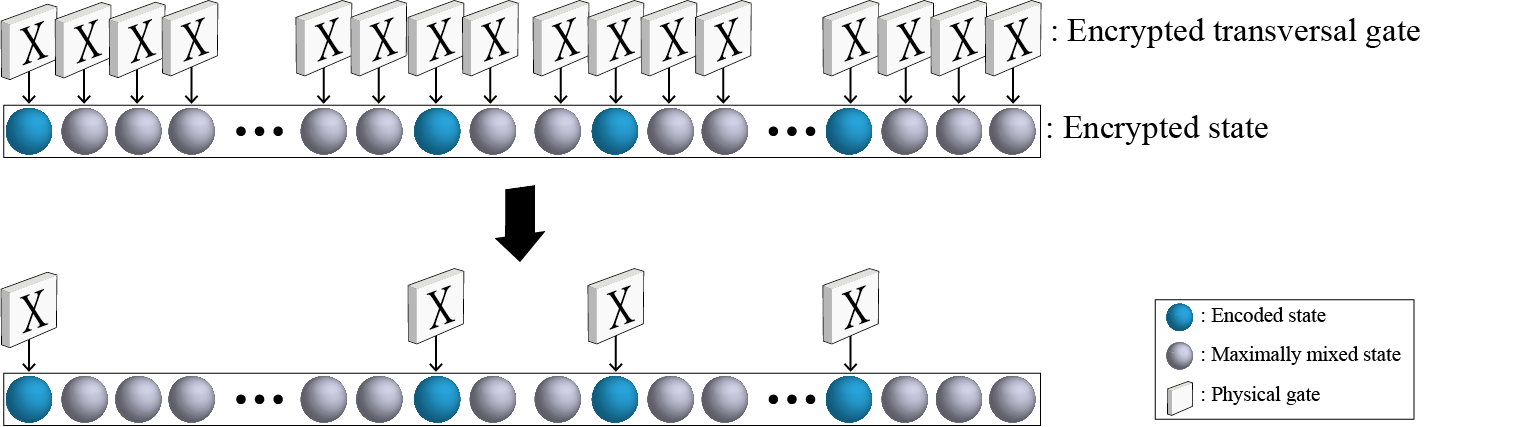}
\caption{\textbf{Transversal gate operation}: An example of performing the logical $X$ operation, which is a transversal gate operation, in the proposed scheme. Performing the same physical gate operation on all qubits  enables transversal gate operations.}
\label{fig:transgate}
\end{figure}
It is well-known that all quantum gates can be decomposed~\cite{dec1, dec2, dec3} or approximated~\cite{app1, app2} using this universal quantum gate set.
Although QECC-encoded data retain QECC characteristics in universal quantum computing, it is essential to investigate the influence of MMSs used for encryption on gate operations and syndrome extraction.
In this section, we discuss the operations involved in the universal gate set of the proposed scheme.
This study focused on the ${H, T, CNOT }$ gate sets within a QECC-encoded environment, even though various forms of universal-gate sets exist~\cite{go99, Ter15}.
Because the QECCs for self-dual and doubly even CSS codes are already constrained, the $H$ and $CNOT$ gates of the universal gate set are implemented transversally.
Furthermore, $S$-gate correction can be transversally executed during gate teleportation to perform non-transversal $T$-gate operations.

The transversal gate operations of the proposed scheme are performed as follows.
In an MMS where the state remains unchanged for all unitary operations, individually applying gate operations to each qubit facilitates transversal gate operations.
Because the logical Pauli and $CNOT$ gates of all CSS codes are transversal~\cite{li17}, they were implemented as shown in Fig.~\ref{fig:transgate}.
The $[[n,k,d]]$ code yields $k$ independent logical Pauli operators and $CNOT$ gates for each code.
However, as the choice of a specific operator can indicate the required decryption, gate operations are performed by multiplying all $k$ independent operators to ensure that the non-transversal gates follow the same principle.
For instance, if the logical $X$ operators for bit-flipping the $i$th data qubit are
denoted as $\bar{X_i}$ in the $[[n,k,d]]$ code, it is sufficient to perform $\bar{X_1}\bar{X_2}...\bar{X_k}$ operations in the proposed scheme, where bit-flip operations are necessary.
This ensures that the server cannot determine the locations of the data bits.

\subsection{Non-transversal $T$-gate operation}
\label{ntg}
\begin{figure}[t]
\centering
\includegraphics[width=.85\linewidth]{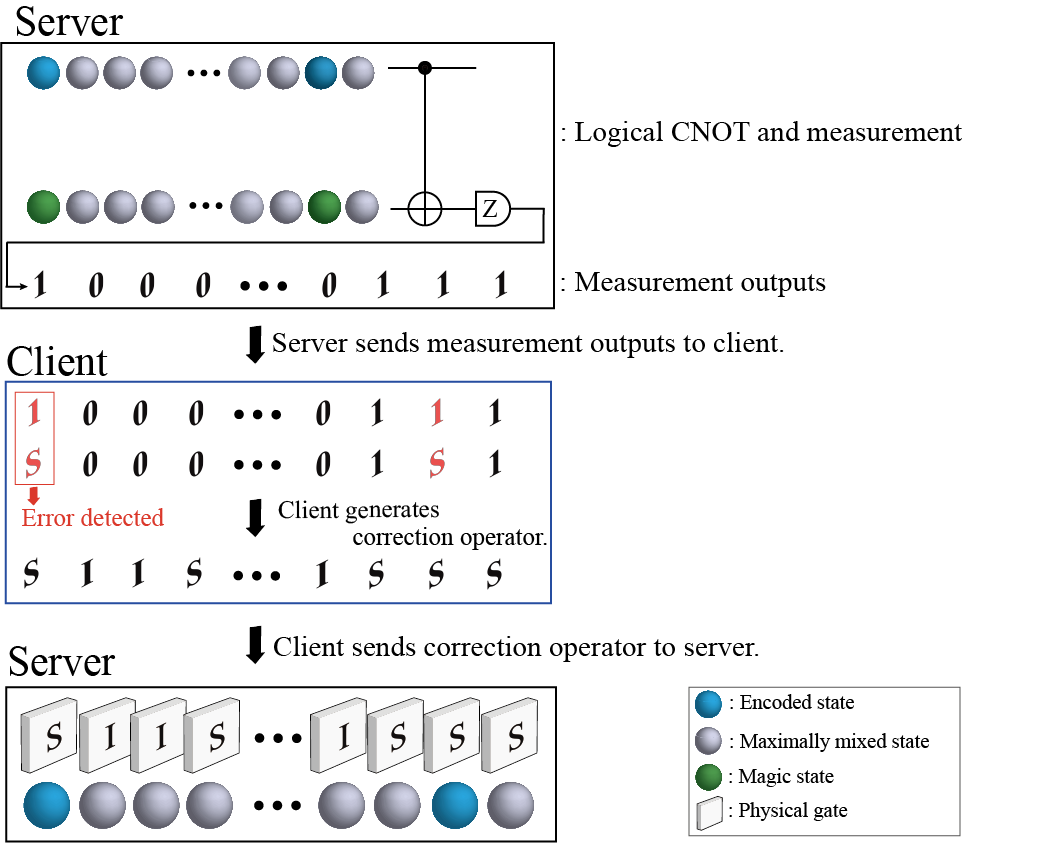}
\caption{\textbf{Non-transversal $T$-gate operation}: An example of performing the $T$-gate operation, which is a non-transversal gate operation, in the proposed scheme. Similar to standard quantum computing, $T$-gate operations can be performed via gate teleportation. In this scenario, after receiving the measurement output from the server, the client detects the occurrence of $S$ errors and sends the encrypted correction operator in response.}
\label{fig:ntransgate}
\end{figure}
For non-transversal $T$-gate operations, the well-known $T$-gate teleportation described in~\cite{ou22} was employed~\cite{go99}, as shown in Fig.~\ref{fig:ntransgate}.
After performing a logical $CNOT$ operation with the magic state, the measurement outputs are sent to the client, which interprets whether the results correspond to logical zeros or ones.

In the method proposed in~\cite{ou22}, the ancillary states (magic, logical-zero, and logical-one) pre-generated by a client are used for deterministic $T$-gate teleportation.
The client does not disclose the specific state at each position on the server and instead, based on the measurement outputs, passes the ancillary-state position to the server for controlled $S$-gate correction, thereby performing encrypted $T$-gate teleportation.
However, this method has several limitations.
First, as the controlled $S$ gate is not transversal in most QECCs, its execution requires decomposition and additional $T$-gates~\cite{VB05, TW18}.
Thus, multiple $T$-gate operations are required to perform a single $T$-gate operation, which increases the computational overhead considerably.
Furthermore, the client may require information regarding how the server stores the states sent by the client.
However, this information may not be ideal in scenarios involving HE~\cite{YH23}.

To circumvent these issues, we employed doubly even CSS codes to create a transversal $S$ gate, and the client performed the following steps after interpreting the measurement results.
In gate teleportation, a logical measurement result of zero indicates the execution of $T$-gate teleportation, whereas a result of one indicates that $S$-gate correction is required~\cite{ou22}.
Therefore, when the result is zero in the proposed scheme, the client instructs the server to apply the following configured operators: identity operators to the data-qubit positions and randomly arranged identity operators and $S$-gates to the remaining MMSs.
However, when the result is one, the client instructs the server to apply an operator composed of $S$-gates to the data qubit position and randomly arranges the identity operators and $S$-gates within the remaining MMSs.
If the operator comprises identity operators and $S$-gates, the server remains oblivious to the logical measurement results and data-qubit positions, regardless of the results.
This approach facilitates non-transversal $T$-gate operations.
The need for client support to interpret measurement outputs presents a problem beyond the inability of the server to perform computing independently, and exposes information about the $T$-gate operations within the circuit being computed by the server, partially compromising circuit privacy.
However, if the server receives numerous encrypted logical qubits for computation, the client cannot determine the qubits subjected to $T$-gate operations.
Moreover, because it does not have information regarding all other transverse gate operations, it cannot reconstruct the entire circuit.

The requirement for client support can be eliminated if the server can interpret the logical measurement results without its assistance.
However, interpreting logical measurement results is equivalent to knowing the data-qubit positions, which poses a challenge that requires further improvements.

We show that the states encoded and encrypted using the proposed scheme enable universal quantum computation by demonstrating a method to perform transversal Pauli operators, $H$ and $CNOT$ gate, and non-transversal $T$-gate within the QHE framework.
Additionally, secure syndrome extraction was implemented to demonstrate the error-correction capabilities of the proposed scheme.

\subsection{Syndrome extraction}
\label{se}
\begin{figure}[t!]
\centering
\includegraphics[width=.85\linewidth]{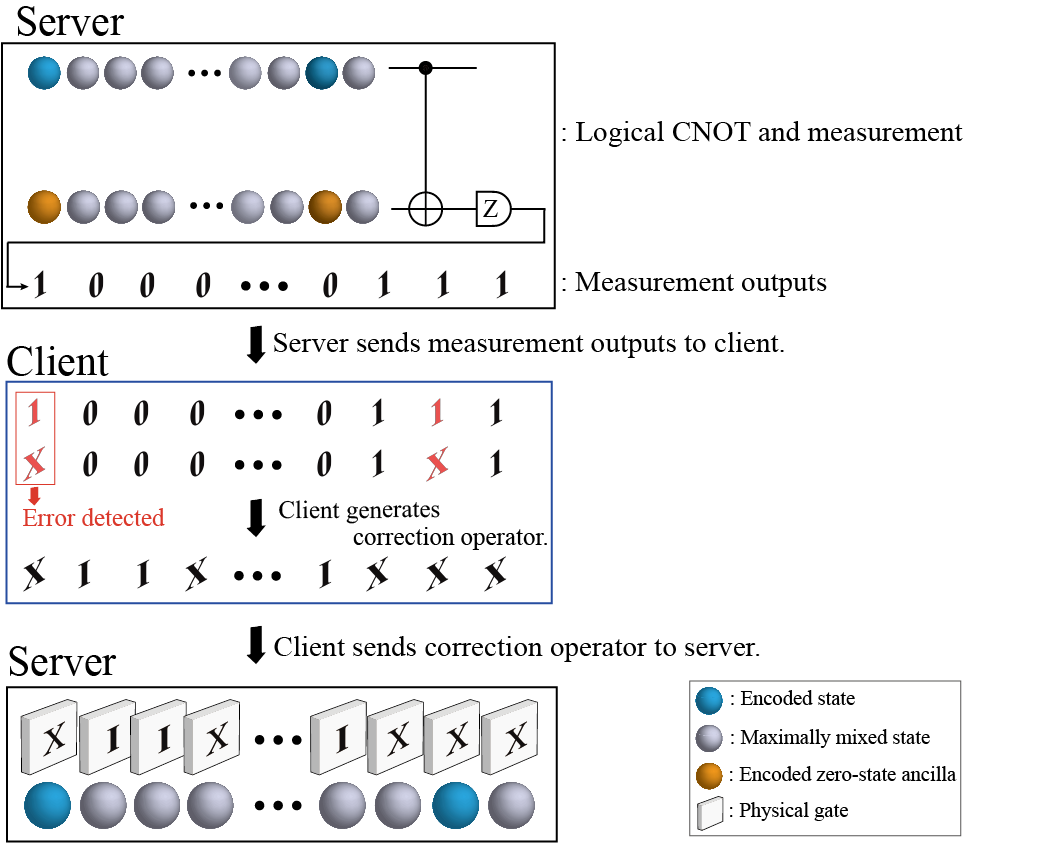}
\caption{\textbf{Syndrome extraction}: Example of the syndrome extraction process using the encoded zero-state ancilla. Similar to non-transversal gate operations, the client responds with the correction operators based on the measurement outputs sent by the server.}
\label{fig:SE}
\end{figure}

Syndrome extraction, which is similar to non-transversal gate operations, requires the processing of measurement results, thereby necessitating client assistance.
The restrictions of QECCs render them usable as CSS code.
To address the ambiguity in the data-qubit positions, the QECCs were constrained to CSS codes that can handle bit- and phase-flip errors independently using the Steane method~\cite{st97}, as shown in Fig.~\ref{fig:SE}.
When using syndrome extraction directly to generate stabilizer QECCs, it is essential to align the operators of the stabilizer generator with the positions of data qubits.
This information can offer a significant clue to the server for determining the data qubit positions, thereby posing a substantial threat to encryption security.
By applying the aforementioned conditions, the syndrome extraction proceeds as follows: based on the Steane method, the server operates on the logical-zero, logical-plus, and encryption states with logical $CNOT$ gates and subsequently measures the ancilla state.
Thereafter, it sends the measurement outputs to the client.
The information in this syndrome pertains only to errors and does not affect security.
Based on this sequence, the client transmits the correction operators, which are the same for each group of $m$-qubit states, to the server.
To analyze the error-correction capability of the encrypted state, we assumed the following depolarizing channel model for communication between the client and the server, as well as for the computational channel of the server:

\begin{equation}
\rho \mapsto \rho'=(1-p)\rho+\frac{p}{3} \left( X\rho X+Y\rho Y+Z\rho Z \right),
\end{equation}
where $\rho$ is the density operator for the channel input and $p$ is the physical error probability.
In this scenario, the logical error probability, $p_L$, of the final state is equivalent to that of standard CSS codes~\cite{pa23} and is expressed as follows:
\begin{equation}
p_L = 1-\sum_{i=0}^t {\binom{n}{i} p^i(1-p)^{n-i}},
\end{equation}
where $t=\lfloor \frac{d-1}{2} \rfloor$ is the error-correction capability.
Despite incorporating MMSs, the logical-error probability is unchanged compared with the standard CSS code because the errors in the depolarizing channel are also unitary.
Additionally, communicating with the client during syndrome extraction allows extracting errors specific to the encoded state.
Thus, the proposed QHE is error-correctable and enables universal quantum computing.

\section{Security proof}
\label{SP}
The quantum error correction code acts as a mechanism to facilitate homomorphic operations.
In this context, the logical operations on the encoded state perform a role analogous to the homomorphic encrypted operations on a homomorphically encrypted state.
Consequently, security must be ensured through other means.
In the proposed scheme, security is derived from how closely the encrypted state approximates the MMSs.
Based on the methods described in~\cite{ou18, ou22}, the security of the proposed method can be assessed by determining the maximal trace distance ($\Delta$) between encrypted states.
Although the gate teleportation is encrypted, revealing the measurement output discloses the quantum state.
Therefore, for security analysis, we must evaluate both the encrypted quantum state and $r$ magic states used throughout the computation.
Thus, the states for security analysis are considered in a form wherein the $\rho$ and $r$ magic states are arranged in rows.
Subsequently, $\Delta$ is defined as follows:
\begin{equation}
\Delta = \max_{\rho_{E||a}, \rho'_{E||a} \in \mathcal{S}} \frac{1}{2} {\Vert \rho_{\textnormal{Eve}} - \rho'_{\textnormal{Eve}} \Vert}_1.
\label{TD}
\end{equation}
where $\Vert \cdot \Vert_1$ is the trace-distance and $\mathcal{S}$ is a set of all possible $\rho_{E||a}$ states.
Let $\rho_{\mathcal{H}}$, $\rho'_{\mathcal{H}}$ denote any two arbitrary encrypted states and $\rho_{Eve}$, $\rho'_{Eve}$ denote any two arbitrary encrypted states perceived by the server or a potential eavesdropper.
The server or eavesdropper perceives these states as a uniformly mixed state of all possible permutation keys, $\kappa \in K$, used for encryption as they lack knowledge of the key.
Subsequently, $\rho_{\textnormal{Eve}}$ is defined as
\begin{equation}
\rho_{\textnormal{Eve}} = \frac{1}{\vert K \vert} \sum_{\kappa \in K}P_{\kappa}\rho_{E||a} P^{\dagger}_{\kappa},
\end{equation}
where $\vert K \vert$ represents the number of possible encryption combinations from the server’s perspective while decrypting without encryption information, that is, the number of all possible combinations of $\kappa$.

With a smaller $\Delta$, it becomes more difficult for the server to distinguish the encrypted messages of the client; hence, less information is available regarding the message in the cipher text.
Therefore, a smaller $\Delta$ implies that the QHE security is enhanced.
By simplifying equation~(\ref{TD}) from the server’s perspective~\cite{ou22}, the maximal trace distance between the encrypted quantum state and MMSs is expressed as
\begin{equation}
\Delta = \sqrt{\frac{2^r}{\vert K \vert}}.
\label{TD2}
\end{equation}
Similar to~\cite{ou22}, the number of $T$-gates, $r$, weakens the security of the proposed scheme, whereas more permutation-key combinations increase it.
This simplified process and the effect of $r$ are explained in the Appendix.

By applying the proposed scheme to equation~(\ref{TD2}), we obtain
\begin{equation}
\Delta_{\textnormal{pro}} = \sqrt{\frac{2^r}{\vert K \vert}}=\sqrt{\frac{2^r}{{{\binom{2m}{1}}}^n}} = \sqrt{\frac{2^r}{(2m)^n}}.
\label{TD3}
\end{equation}
This indicates that the security and error-correction capabilities of the proposed scheme improve simultaneously as $n$ increases.
Thus, the proposed scheme provides exponentially increasing security as $n$ increases, thereby ensuring a constant security level by adjusting $n$ based on $r$.

\section{Performance analysis}
In this section, we analyze the security of the proposed scheme by comparing its $\Delta_{\textnormal{pro}}$ with $\Delta_{\textnormal{pre}}$ of the permutation-key-based QHE scheme employed in~\cite{ou22}, as well as their error-correction capabilities when both schemes use the same overhead.
By applying equation~(\ref{TD2}) to the previous permutation-key-based QHE scheme, $\Delta_{\textnormal{pre}}$ is computed as follows:
\begin{equation}
\Delta_{\textnormal{pre}}=\sqrt{\frac{2^r}{{\binom{2m}{m}}}} \sim \sqrt{\pi n\frac{2^r}{4^{m}}},
\label{dpre1}
\end{equation}
The approximation in equation~(\ref{dpre1}) is obtained using Stirling’s formula~\cite{JD91}.
Hence, to exponentially enhance security, the previous permutation-key-based QHE scheme requires increasing the number of MMSs, whereas the proposed scheme requires extending the QECC lengths.
Therefore, in a scenario wherein the same number of qubits is used and an equivalent level of security is required, the conventional permutation-key-based QHE scheme cannot employ longer QECCs than the proposed scheme.
Additionally, even if the value of $n$ increases, that of $k$ remains fixed.
In such cases, the efficient construction of QECCs can increase the minimum distance associated with the error-correction capability~\cite{ct06, ct08}.
Additionally, the proposed scheme restricts the encryption to single-qubit messages to mitigate the risk of information leakage owing to differences in the logical operators of messages encrypted with more than two qubits.
Thus, higher $n$ values can enhance the error-correction capability of the proposed scheme.

To compare the securities of the conventional permutation-key-based QHE and proposed schemes, inequalities were established based on the $\Delta$ values of both schemes, yielding
\begin{equation}
\label{pa}
\frac{\pi n}{4^{m}} > \frac{1}{4^{m}} >\frac{1}{(2m)^n}.
\end{equation}
The term $\pi n$ in the numerator of $\Delta_{\textnormal{pre}}$ weakens security; however, to simplify the comparison, we eliminated it.
The inequality in equation~(\ref{pa}) solved with $n$ is expressed as
\begin{equation}
n > \frac{\log{4^{m}}}{\log{2m}}.
\label{ine}
\end{equation}
In the integer domain where $m\leq n$, the proposed scheme exhibits higher security.
Even if $m$ is larger, it must be more than twice the size to ensure that the conventional scheme offers higher security.
As a simple example, when using an equal number of qubits ($n\times 2n$ in total), $\Delta_{\textnormal{pro}}=\sqrt{\frac{2^r}{(2n)^n}}$, and $\Delta_{\textnormal{pre}} \sim \sqrt{\frac{2^r}{4^{n}}}$.
Thus, the proposed scheme offers higher security under the same overhead, because the growth rate of $(2n)^n$ is faster than that of  $4^{n}$ for $n>2$.

\section{Conclusion}
This paper presented an efficient QHE scheme based on CSS codes that supports error correction.
Considering the current development of various quantum computers that support cloud computing and the essential role of QECCs in achieving future fault-tolerant universal quantum computing, efficiently using error correction and HE is crucial.
The proposed scheme enhanced efficiency by integrating encryption and encoding into a single process, reducing resource overhead while simultaneously improving security and error-correction capabilities as the code length $n$ increases.
Our analysis showed that the scheme offers higher security compared to previous permutation-key-based QHE methods, especially when the number of maximally mixed states $m$ is not more than twice the QECC length.
Furthermore, we simplified non-transversal gate operations by eliminating the need for a secure pre-generated set of ancilla qubits and presented a novel approach for syndrome extraction.
Our QHE scheme, which efficiently realizes the quantum homomorphic encryption, provides a new insight into the field of quantum cryptography.
It is also expected to be potentially applied to a secure quantum cloud computing which is the one of the imminent challenges in this field.

\backmatter

\bmhead{Acknowledgements}
This research was supported by Korea Institute of Science and Technology Information (KISTI). (No. K25L5M2C2). This research was supported by the National Research Council of Science \& Technology (NST) grant by the Korea government (MSIT) (No. CAP22053-000)

%%\bigskip
%%\begin{flushleft}%
%%Editorial Policies for:

%%\bigskip\noindent
%%Springer journals and proceedings: \url{https://www.springer.com/gp/editorial-policies}

%%\bigskip\noindent
%%Nature Portfolio journals: \url{https://www.nature.com/nature-research/editorial-policies}

%%\bigskip\noindent
%%\textit{Scientific Reports}: \url{https://www.nature.com/srep/journal-policies/editorial-policies}

%%\bigskip\noindent
%%BMC journals: \url{https://www.biomedcentral.com/getpublished/editorial-policies}
%%\end{flushleft}

\begin{appendices}
\section{Derivation of security proof}
\label{ap}
The trace norm of equation~(\ref{TD}) can be expressed,
\begin{eqnarray}
\vert \rho_{\textnormal{Eve}} - \rho'_{Eve} \vert_1&=\mathrm{Tr}[\sqrt{(\rho_{Eve} - \rho'_{Eve})^2}] \nonumber \\
                                        &=\mathrm{Tr}[M(\rho_{Eve} - \rho'_{Eve})],
\end{eqnarray}
where $M=\sum_{i}\mathrm{sgn}(\lambda_{i})|v_i \rangle\langle v_i|$ and $\sum_{i}\lambda_{i}|v_i \rangle\langle v_i|$ is the spectral decomposition of $\rho_{Eve} - \rho'_{Eve}$ ($\mathrm{sgn}(x)$ is the signum function groups zero with the positive numbers).
We define $\frac{1}{\vert K \vert}\sum_{\kappa \in K}P_{\kappa}MP^{\dagger}_{\kappa}$ as $\tilde{M}$.
Using the cyclic property of the trace and non-negativity of the trace norm, we get
\begin{equation}
\Vert \rho_{Eve} - \rho'_{Eve} \vert_1=\vert\mathrm{Tr} \tilde{M}(\rho_{E||a} - \rho'_{E||a})\vert.
\label{TDP1}
\end{equation}
Following the approach in \cite{ou18, ou22}, we decompose $\tilde{M}$ and $\rho_{E||a} - \rho'_{E||a}$ individually using Pauli decomposition, equation~(\ref{TDP1}) can be expressed as
\begin{equation}
\Vert \rho_{Eve} - \rho'_{Eve} \Vert_1=\vert \mathrm{Tr}\sum_{A \in \mathcal{S}}{a_{A}\tilde{\sigma}_{A}}
\sum_{\mathrm{v} \in \Omega}\frac{r_\mathrm{v}-r'_\mathrm{v}}{2^{2(r+1)nm}}\sigma_{\varphi(\mathrm{v})} \vert,
\end{equation}
where $A$ is the arbitrary $(r+1)$-by-$2mn$ matrix with each element belonging to $\mathbb{Z}_4$.
The matrix $\sigma_A$ is constructed by placing Pauli operators at each position according to the values of the elements of $A$.
Furthermore, $\tilde{\sigma}_{A}$ is the sum of $\sigma_A$ over all possible permutation-keys $P_{\kappa}$.
$\Omega$ is the set of all nonzero column vectors with length $(r+1)$ and each element belonging to $\mathbb{Z}_4$.
$\varphi(\mathrm{v})$ is also an $(r+1)$-by-$2mn$ matrix for $\mathrm{v} \in \Omega$.
Its first $n$ columns are identical to $\mathrm{v}$ and the remaining $2mn-n$ columns are all zeros.
$\mathcal{S}$ is some minimal subset of the set of all matrices $\mathcal{M}$ with $(r+1)$ rows and $2n$ columns and entries from $\{0,1,2,3\}$ satisfying $\{\tilde{\sigma}_{A}:A\in\mathcal{S}\}=\{\tilde{\sigma}_{A}:A\in\mathcal{M}\}$.
Now if $\mathrm{v}$ in $\Omega$, then $\varphi(\mathrm{v})$ in $\mathcal{S}$.
$a_A$, $r_\mathrm{v}$, $r'_\mathrm{v}$ are all real constants associated with the Pauli decomposition of each term.
After, simplifying the above using orthogonality, we get
\begin{equation}
\vert \rho_{Eve} - \rho'_{Eve} \vert_1=\vert {\sum_{\mathrm{v} \in \Omega}{a_{\varphi(\mathrm{v})}(r_\mathrm{v}-r'_\mathrm{v})}}\vert,
\label{TDP2}
\end{equation}
Applying the Cauchy-Schwarz inequality to the aforementioned yields
\begin{equation}
\vert \rho_{Eve} - \rho'_{Eve} \vert_1 \leq \sqrt{ \sum_{\mathrm{v} \in \Omega}{\left( a_{\varphi(\mathrm{v})} \right)^2}} \sqrt{\sum_{\mathrm{v} \in \Omega}(r_\mathrm{v}-r'_\mathrm{v})^2}.
\label{TDP3}
\end{equation}
To find the upper bound of the first term on the right-hand side of the inequality, we obtain upper and lower bounds on $\mathrm{Tr}(\tilde{M}^2)$.
By Hölder’s inequality, we get
\begin{equation}
\mathrm{Tr}(\tilde{M}^2) \leq \vert \tilde{M} \vert_1 \vert \tilde{M} \vert_\infty = 2^{2(r+1)nm},
\end{equation}
where $\vert \cdot \vert_\infty$ is the $\infty$-norm, because $\tilde{M}$ is a matrix related to $(r+1)2mn$-qubits, so $\vert \tilde{M} \vert_1$ is $2^{2(r+1)nm}$.

The lower bound on $\mathrm{Tr}(\tilde{M}^2)$ is
\begin{eqnarray}
\mathrm{Tr}(\tilde{M}^2) &= \sum_{A\in \mathcal{S}}a^2_{A}\mathrm{Tr}(\tilde{\sigma}^2_{A}) \nonumber \\
                         &\geq \sum_{\mathrm{v} \in \Omega}a^2_{\varphi(\mathrm{v})}\mathrm{Tr}(\tilde{\sigma}^2_{\varphi(\mathrm{v})}),
\end{eqnarray}
Since the set containing $\varphi(\mathrm{v})$ is smaller than $S$, the above inequality holds.
The $\tilde{\sigma}_{\varphi(\mathrm{v})}$ represents the sum of $\vert K \vert$ variations of $\sigma_{\varphi(\mathrm{v})}$ and all Pauli operators in $\tilde{\sigma}^2_{\varphi(\mathrm{v})}$ become the Identity, thus $\mathrm{Tr}(\tilde{\sigma}^2_{\varphi(\mathrm{v})})= \vert K \vert 2^{2(r+1)nm}$.
Thus
\begin{equation}
\mathrm{Tr}(\tilde{M}^2) \geq \sum_{\mathrm{v} \in \Omega}a^2_{\varphi(\mathrm{v})}\vert K \vert 2^{2(r+1)nm}.
\end{equation}
Thus, the first term on the right-hand side of equation~(\ref{TDP3}) is transformed to
\begin{equation}
\sqrt{ \sum_{\mathrm{v} \in \Omega}{\left( a_{\varphi(\mathrm{v})} \right)^2}} \leq \frac{1}{\sqrt{\vert K \vert}}.
\end{equation}

To determine the upper bound of the second term $\sqrt{\sum_{\mathrm{v} \in \Omega}(r_\mathrm{v}-r'_\mathrm{v})^2}$ on the right-hand side of equation~(\ref{TDP3}), we decompose the arbitrary quantum message state $\rho_m$ using Pauli decomposition to be,
\begin{equation}
\rho_m = \frac{I^{\otimes(r+1)}+\sum_{\mathrm{v} \in \Omega} r_\mathrm{v} \sigma_\mathrm{v}}{2^{(r+1)}}.
\label{qs}
\end{equation}
The maximum eigenvalue of this state is $2^{-h}$, where $h$ is their minimum entropies~\cite{ou18}.

Using Hölder’s inequality, we can obtain an upper bound,
\begin{equation}
\mathrm{Tr}[(\rho_m-\rho'_m)^2] \leq \vert \rho_m-\rho'_m \vert_1 \vert \rho_m-\rho'_m\vert_\infty.
\label{sup}
\end{equation}
By utilizing the triangular inequality, $\vert A-B \vert \leq \vert A \vert+\vert B \vert$, we can derive,
\begin{eqnarray}
\vert \rho_m-\rho'_m \vert_1  &\leq \vert \rho_m \vert_1 +\vert \rho'_m \vert_1 \leq 2,\\
\vert \rho_m-\rho'_m \vert_\infty &\leq \vert \rho_m \vert_\infty+\vert \rho'_m \vert_\infty = 2^{-h+1}.
\end{eqnarray}
Therefore, substituting the results of the above equations into equation~(\ref{sup}), we obtain $\mathrm{Tr}[(\rho_m-\rho'_m)^2] \leq 2^{-h+2}$.

Subsequently, rearranging $\mathrm{Tr}[(\rho_m-\rho'_m)^2]$ using orthogonality of Pauli operators in equation~(\ref{qs}),
\begin{eqnarray}
\mathrm{Tr}[(\rho_m-\rho'_m)^2] &= \mathrm{Tr}[\sum_{\mathrm{v} \in \Omega}(r_\mathrm{v}-r'_\mathrm{v})^2 2^{-2(r+1)}I^{\otimes (r+1)}] \nonumber \\
                                &= \sum_{\mathrm{v} \in \Omega} (r_\mathrm{v}-r'_\mathrm{v})^2 2^{-(r+1)}.
\end{eqnarray}
Therefore the upper bound of $\sqrt{\sum_{\mathrm{v} \in \Omega}(r_\mathrm{v}-r'_\mathrm{v})^2}$ is
\begin{equation}
\sqrt{\sum_{\mathrm{v} \in \Omega}(r_\mathrm{v}-r'_\mathrm{v})^2} \leq 2\sqrt{2^{(r+1)-h}}.
\end{equation}
Therefore, if we denote the $h=0$ and omit the influence of the message qubit, the maximal trace-distance between encrypted states can be simplified as follows,
\begin{eqnarray}
\Delta &= \max_{\rho_{E||a}, \rho'_{E||a} \in \mathcal{S}} \frac{1}{2} \vert \rho_{Eve} - \rho'_{Eve} \vert_1 \nonumber \\
       &= \sqrt{\frac{2^r}{\vert K \vert}}.
\end{eqnarray}

\section{Performance analysis based on the combination of $n$ and $m$}
\begin{figure}[ht!]
\centering
\includegraphics[width=.55\linewidth]{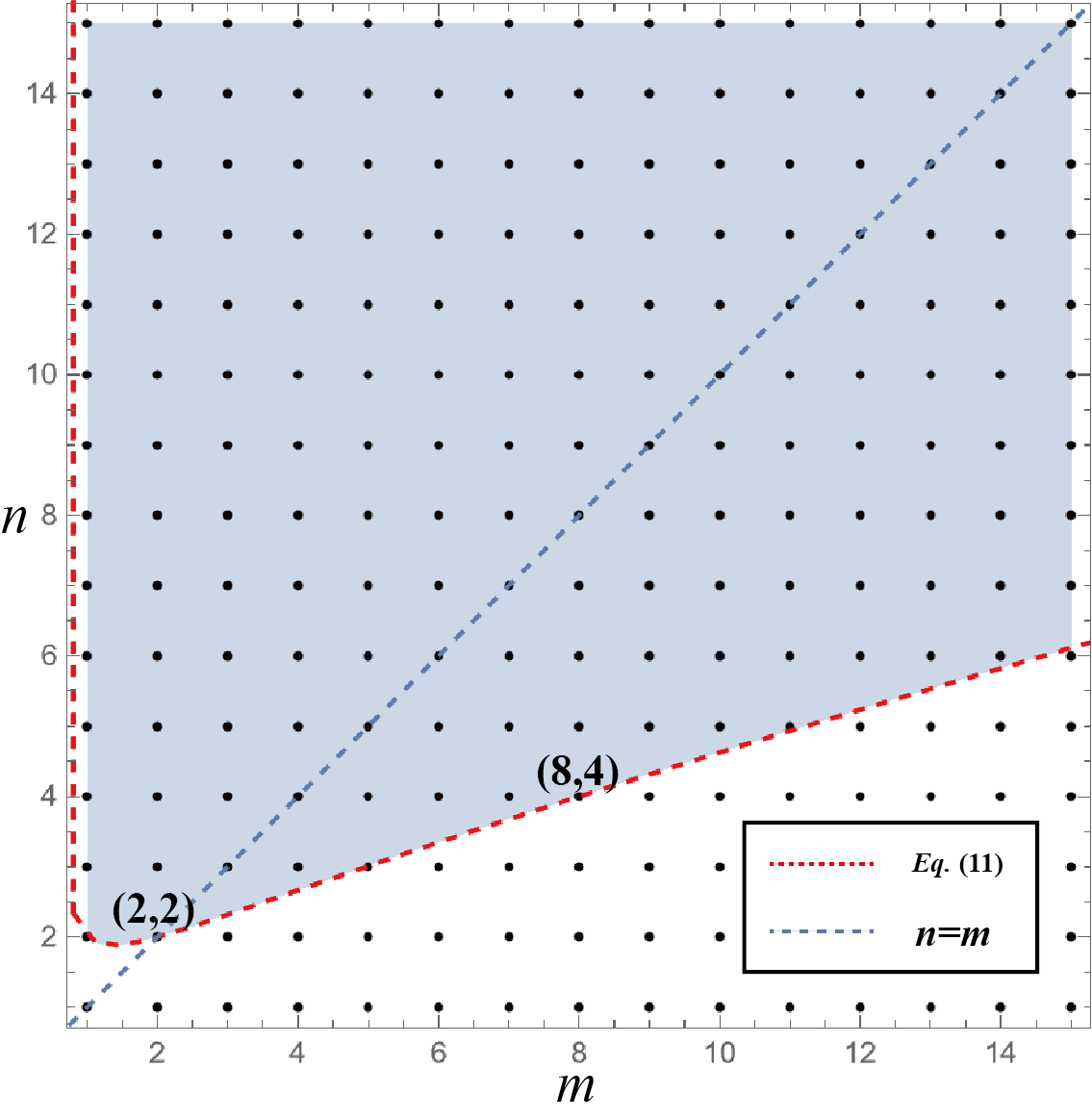}
\caption{\textbf{Security comparison of the proposed and conventional QHE schemes}: The graph shows the region wherein the proposed scheme exhibits higher security according to equation~(\ref{ine}).
The blue-shaded area represents the higher security region of the proposed scheme, whereas the black dots represent the actual usable integer values of $m$ and $n$ for each scheme.
When $n>m$, the proposed scheme demonstrates higher security.
The lower bound in equation~(\ref{ine}) passes through points $(2,2)$ and $(8,4)$, and to simplify the slope comparison, we have included the m=n graph of the blue dashed line.}
\label{fig:gr}
\end{figure}
The graph in Fig.~\ref{fig:gr} corresponds to equation~(\ref{ine}), which demonstrates that, in the integer domain where $m\leq n$, the proposed scheme exhibits higher security.
Additionally, the slope of equation~(\ref{ine}), $m>n$ and is consistently less than one and asymptotically approaches zero as $m$ approaches infinity.
Therefore, in the region where $m>n$, there is always an area between lines $m = n$ and equation~(\ref{ine}) where the proposed scheme exhibits higher security.
Evidently, it outperforms the conventional permutation-key-based QHE scheme for most feasible combinations of $m$ and $n$.
By numerically comparing the areas, the proportion of the region where the proposed scheme exhibits better security was determined to be 66.17\% when $1 \leq n,m \leq 5$; 84.10\% when $1 \leq n,m \leq 50$; and 98.34\% when $1\leq n,m \leq 5000$.


\end{appendices}


\end{document}